\documentclass[aps,prd,amsmath,showpacs,preprint]{revtex4}
\usepackage{amsmath}
\usepackage[dvips]{graphicx}
\usepackage[cp1251]{inputenc}
\usepackage[english]{babel}

\newcommand{\be}{\begin{equation}}
\newcommand{\ee}{\end{equation}}
\newcommand{\bn}{\begin{eqnarray}}
\newcommand{\en}{\end{eqnarray}}
\newcommand{\bd}{\begin{displaymath}}
\newcommand{\ed}{\end{displaymath}}
\newcommand{\bnn}{\begin{eqnarray*}}
\newcommand{\enn}{\end{eqnarray*}}
\def\Ref#1{(\ref{#1})}
\def\Journal#1#2#3#4#5#6{#1, \ "#2," \ #3 \ {\bf #4}, \ #5 \ (#6)}
\def\journal#1#2#3#4#5#6{#1, \ "#2," \ #3 \ {\bf #4}, \ #5 \ (#6)}

\begin{document}
\inputencoding{cp1251}
\title{Non-Markov stochastic dynamics of real epidemic process of respiratory infections}
\author{Renat Yulmetyev} \email{rmy@dtp.ksu.ras.ru}
\author{Natalya Emelyanova}\author{Sergey Demin}
\author{Fail Gafarov} \affiliation{Department of Physics,
Kazan State Pedagogical University,  Mezhlauk Street 1, 420021
Kazan,Russia}
\author{Peter H\"anggi}\affiliation{Department of
Physics, University of Augsburg, Universit\"atsstrasse 1,D-86135
Augsburg,Germany}
\author{Dinara Yulmetyeva (M.Dr.)}\affiliation{Division of
Therapy, Republican Clinical Hospital,  Orenburgskii Trakt 79,
420112 Kazan, Russia}
\begin{abstract}
The study of social networks and especially of the stochastic
dynamics of the diseases spread in human population has recently
attracted considerable attention in statistical physics. In this
work we present a new statistical method of  analyzing  the spread
of epidemic processes of grippe and acute respiratory track
infections (ARTI) by means of the  theory of discrete non-Markov
stochastic processes. We use the results of our last theory (Phys.
Rev. E 65, 046107 (2002)) to study statistical memory effects,
long - range correlation and discreteness in real data series,
describing the epidemic dynamics of human ARTI infections and
grippe. We have carried out the comparative analysis of the data
of the two infections (grippe and ARTI) in one of the industrial
districts of Kazan, one of the largest cities of Russia. The
experimental data are analyzed by the power spectra of the initial
time correlation function and the memory functions of junior
orders, the phase portraits of the four first dynamic variables,
the three first points of the statistical non-Markov parameter and
the locally averaged kinetic and relaxation parameters. The
received results give an opportunity to provide strict
quantitative description of the regular and stochastic components
in epidemic dynamics of social networks taking into account their
time discreteness and effects of statistical memory. They also
allow to reveal the degree of randomness and predictability of the
real epidemic process in the specific social network.
\end{abstract}
 \maketitle

{\bf   The functioning of network-organized statistical systems
essentially depends on the  nature of interaction between their
elements.  It is especially due to the effect of disease-causing
contacts and topology of networks. This reason as well as  the
variety of the displayed nonlinear behavior have made this problem
the subject of several studies by  fundamental methods of
statistical physics. To  analyse the epidemic and disease dynamics
complexity, it is necessary to understand the basic principles and
notions of  its distribution in long-time memory social media.
Here we consider the problem from a theoretical and practical
viewpoint and present the quantitative evidence confirming the
existence of stochastic long-range memory and robust chaos in a
real time series of respiratory  infections of human upper
respiratory track. We also discuss the implications of discrete
non-Markov stochastic processes in  real social complex systems
from the point of view of the recent theory (Phys. Rev. E 65,
046107 (2002)).}

\section {Introduction}
In the last few years it has been well recognized that the study
of epidemic and disease dynamics in social networks is a relevant
theoretical issue concerning the spreading of  viruses and
diseases of various nature \cite{New, Pas, Grass, Amar}. It is
connected with the phase transitions in models of agent spreading
in nonsteady systems. In particular, it concerns the development
of the models describing an epidemic spread, forest fires, growth
of populations, activity of catalyzers, formation of stars and
galaxies. The given problem might be of interest for all those
interested in physical problems, related to the critical
phenomenon in the universality class of Reggeon field theory,
contact processes, ordinary and directed percolation and
scale-free properties in many real systems such as Internet, World
Wide Web, food webs, protein and neural networks.

This study is of significant interest for modern biomedicine,
ecology and economy \cite{Burk}-\cite{Snac}. The epidemic means
that a large number of people in a certain country simultaneously
develop a disease  at a certain period of time. In  case of
"pandemic" people in several countries develop the same disease at
the same period of time. The first cases of respiratory infections
with fatal outcome in the USA and Europe were described at the
beginning of the 20th century. About 20 million people died then
of  "Spanish fever" within the period of 10 months. The pandemic
of "Asian" and "Hongkong" grippe with smaller quantity of fatal
outcomes \cite {Burk}-\cite {Stuart} were observed in the middle
of the 20th century. The period of pandemic processes is about
30-40 years. As a rule, epidemics break out in the autumn or in
the winter in Northern Hemisphere and in the spring or in the
summer in Southern Hemisphere. The duration of epidemics is about
1-3 months. Seasonality is one of the typical demonstrations of
the phase development of epidemic processes. Respiratory
infections affect people of any age. However children aged 1-14
develop the disease four times more often than adults. Each
epidemic process of respiratory infections is accompanied by the
increasing death rate. The parameters of grippe death rate in the
world is $0,01-0,2 \%$. The greatest number of deaths caused by
grippe is connected with serious complications after the
infection. Children (up to 2 years old)  and elderly people (over
65 years old)   are more liable to this infection and die more
often.

Various authors have discovered that the number of people
developing the disease depends to a great extent on ecological and
economic conditions. The most common diseases the population of
the earth suffers from are infections of respiratory organs among
which grippe and acute infections of the upper respiratory track
are main components. The low standard of living as well as the
pollution of the environment have had their negative influence on
people's health. This is the main reason for the worsening
epidemic situation in the world. Respiratory infections cause
significant economic damage. Expenses caused by epidemics affect
successful development of economy due to the expenses needed for
financing scientific actions to prevent epidemics. These expenses
are commensurable only with the sums earmarked to prevent heart
diseases \cite {Peng1}-\cite{Stan6}. In many countries of the
world great money is spent on the development of vaccines and
serums, as well as prevention of epidemic outbreaks. However the
viruses of grippe and some other respiratory infections have great
variability. There is a great variety of grippe virus cultures.
Therefore the coming grippe variant is difficult to predict. The
identification of the virus presents a great problem, consequently
the prevention of epidemic outbreaks is complicated. Thus new
preventive measures and methods of predicting possible epidemic
outbreaks \cite {Web}-\cite {Snac} are of great interest. The
unexpected outbreak of grippe in March 2003 is a good example of
the case.

Now various methods of description and prediction of epidemic and
pandemic processes are developed in medicine \cite {Web}-\cite
{Snac}. Hypotheses about risk factors have been formulated as a
result of these  studies. The standard statistics indices \cite
{Snac} are used in traditional epidemic studies. The basic data
yield the number of diseases, their intensive parameters and
average sizes. The extensive parameters, the cumulative data, the
relative number of presentation, the ratio parameters and
standardized parameters are used depending on the specific
features of the epidemic process. However existing methods are not
efficient enough. A model for the spread of an infection is
analyzed \cite{Kuper} for different population structures. For
more ordered systems, there exists fluctuating endemic state of
low infection. At a finite value of the disorder of the network, a
transition to self-sustained oscillations in the size of the
infected subpopulation has been recorded. A spatial model related
to bond percolation for the spread of a disease that includes
variation in the susceptibility to infection has also been
 considered in Ref. \cite {Warren}.

Methods of networks are all important for fundamental solution of
the problems related to the spread of epidemics of grippe  and
ARTI. Traditionally these systems have been modelled as random
graphs, however topology and evolution of real networks are
governed by robust organization principles. In this connection it
should be mentioned that an excellent review of complex networks
in a wide range of systems in wild nature and human society was
made in Ref. \cite {Alb}. A method for embedding scale-free
networks in Euclidean lattices was proposed in Ref. \cite{Rozen}.
The critical dynamics, Glauber type and Kawasaki type, on two
typical small-world networks, adding type and rewriting type, was
studied in Ref. \cite{Zhu}. Newman \cite{Newman} considered
assortative mixing of various types using empirical network data,
analytic models, and numerical simulation. He showed that
assortative (or disassortative) mixing is indeed present in many
networks, it can be measured, and its effect on network structure
and behavior can be examined. Generalized the Bethe-Peierls
approach to random networks with degree correlations, and the
analysis of the vertex cover (VC) problem as a prototype
optimization problems defined over graphs \cite{Vaz} has shown
that uncorrelated power-law networks are simple from the point of
view of combinatorial optimization, and inhomogeneities of
neighboring vertices can be exploited.

In hierarchical networks, the degree of clustering characterizing
different groups follows a strict scaling law, which can be used
to identify the presence of hierarchical organization in real
networks \cite{Rav}. The consideration of structured scale
free-networks restores the order-disorder transitions in spite of
the hubs, but the value of the order parameter for the disordered
state reveals the existence of ordered clusters \cite{Klemm}. A
nonstandard, susceptible-exposed-infected (SEI), compartmental
model for disease epidemics comprising latency and temporal decay
in the rates of infection, also known as quenching, has been
examined in Ref. \cite{Filipe}. Eventually, one can recognize that
the study and solutions of epidemic models has been of a
long-standing interest in theoretical physicists motivated by
analogy between spreading phenomena in epidemic and physical
systems and by application of analytical techniques.

Numerous methods are successfully used in  statistical physics to
describe the distinctive characteristics  of chaotic dynamics of
various networks. However, three vexing features that are
difficult for the detailed analysis  can be observed in real
networks. Among them: nonstationarity, nonlinearity, and
nonequilibrium phenomena. Furthermore, the significant
peculiarities of networks are directly related to the discreteness
in time of object-subject registration response. Non-Markov and
long-range statistical memory effects also play the leading part
in epidemic dynamics in social networks. For this reason  we offer
a new statistical approach in the research of epidemic processes
with the help of the our statistical theory of non-Markov
stochastic discrete processes \cite {Yulm1, Yulm2, Yulm3} in this
work. This method has already been used by the authors to study
various real problems in cardiology \cite {Yulm1, Yulm3},
seismology \cite {Yulm2}, neuropsychology \cite {Yulm4, Yulm5} and
neurophysiology \cite {Yulm6}. With the help of this method we
intend to find the full spectrum of the dynamic kinetic and
spectral characteristics of the studied complex system. We shall
receive a more detailed representation of the stochastic dynamics
of epidemic processes taking into account the discrete behavior
and nonstationarity of the dynamic and kinetic epidemic
parameters.

This work deals with the study of statistical properties of
epidemic dynamics. The key distinctions of grippe  and respiratory
infections epidemic processes are defined with the help of the
long-range memory effects and statistical effects of non-Markovity
in these processes. In particular, one of the key problems
consists in revealing the role of randomness, regularity and
predictability of epidemic processes.

The paper has the following structure. In Section 2 we show the
basic points of our statistical theory of nonstationary discrete
non-Markov processes in complex systems \cite {Yulm1, Yulm2,
Yulm3}. The description of the experimental data and the technique
of their processing are given in Section 3. The received results
and their analysis are submitted in Section 4. The conclusions are
submitted in Section 5.

\section{Statistical theory of
nonstationary discrete non-Markov processes  in complex systems.
Basic concepts and definition}

A brief description  of the theory is presented here. We will
consider a spread of epidemic as a time evolution of a patient's
number (discrete random variable $x (T)$). Therefore we can use
results of the statistical theory of discrete non-Markov
non-stationary processes for complex systems. The developed
description of the theory and the representation of quantities
used here can be found in the works published earlier
\cite{Yulm1}-\cite{Yulm6}.

While analyzing complex systems we obtain discrete equidistant
series of experimental data, the so-called random variable \be X =
\{ x (T), x (T +\tau), x (T+2\tau), \cdots, x (T+k\tau), \cdots, x
(T +\tau N-\tau) \}. \label {f5} \ee It corresponds to a time
series of measured signal during the time $(N-1) \tau $, where $
\tau $ is discretization time  of the signal. In this work  we
take the number of patients per day $x (T)$  as a measurable
parameter, $(N-1)\tau$ is a time interval of data recording (6
years), $ \tau $  is one day.

For the dynamical analysis, it is more convenient  to use a
normalized time correlation function (TCF).  For nonstationary
discrete processes TCF has the following form  ($t=m\tau, N-1 > m
> 1$)
\be a(t)=\frac{1}{(N-m) \sigma(0)\sigma(t)}
\sum_{j=0}^{N-1-m}\delta x(T+j\tau)\delta x(T+(j+m)\tau), \label
{f6} \ee where $\sigma(0)$ and $\sigma(t)$ is the variances of the
initial $(t=0)$ and final (at moment $t$) dynamic states of the
systems, correspondingly. The properties of TCF $a (t) $ are
determined by the conditions of normalization (at t=0) and
attenuation of correlations (at  t ${\to \infty}$) \be \lim_{t \to
0} a(t)=1, \lim_{t \to \infty} a(t)=0. \label {f7} \ee If we take
into account the non-stationarity and discreteness  of complex
systems for real processes, the kinetic equation for TCF $a (t)$
has the form of a closed set of the finite-difference kinetic
equations of the non-Markov type \cite{Yulm1,Yulm2,Yulm3,Yulm4}:
\be \frac {\Delta a (t)} {\Delta t} = \lambda_1 a (t) -\tau
\Lambda_1 \sum _ {j=0} ^ {m-1} M_1 (j\tau) a (t-j\tau) .\label
{f8} \ee Here $ \Lambda_1 $ is a relaxation parameter with the
dimension of square of frequency, and parameter $\lambda_1$
describes the eigen-spectrum of Liouville's quasioperator $ \hat L
$ \be \lambda_1=i\frac {< {\bf A} _k^0 (0) \hat L
 {\bf A}_k^0 (0) >} {< | {\bf A} _k^0 (0) | ^2 >}, ~~ \Lambda_1 =\frac {< {\bf
A}_k^0 (0) \hat L _ {12} \hat L _ {21} {\bf A} _k^0 (0) >} {< |
{\bf A}_k^0 (0) |^2 >}, \label {f9} \ee where angular brackets
mean a scalar product of state vectors.

The function $M_1 (j\tau) $ in the rhs of Eq. \Ref {f8} represents
the first order memory function \be M_1(j \tau)= \frac{< {\bf
A}_k^0 (0) \hat L_{12} \{ 1+i\tau \hat L _ {22} \} ^j \hat L _
{21} {\bf A} _k^0 (0) >} {< {\bf A}_k^0 (0) \hat L _ {12} \hat L _
{21} {\bf A} _k^0 (0) >}, ~~ M_1 (0) =1.\label {f10} \ee

In Eqs. \Ref {f9} and \Ref {f10} operator $\hat L$ is a
finite-difference operator \bn i \hat L =
\frac{\bigtriangleup}{\bigtriangleup t}, \ \ \bigtriangleup t =
\tau, \nonumber \en where $\tau$ is a discretization time  step, $
\hat L_{ij}= \Pi_{i} \hat L \Pi_{j}$ ( $i, j=1,2$) are matrix
elements of splittable Liouville's quasioperator,  $ \Pi_{1}=\Pi,
\Pi_{2}= P = 1-\Pi$ and $\Pi$ are projection operators (for more
details, see, Refs. \cite{Yulm1}-\cite{Yulm5}). It is easy to see,
that in Eq. \Ref {f10} we deal with the time correlation of new
orthogonal dynamic variable $ \hat L _ {21} {\bf A} _k^0 (0)$.

Eq. \Ref {f8} represents the first equation in the chain of
finite-difference kinetic equations with memory for the discrete
TCF $a (t)$. One can recognize that the memory function $M_1 (t) $
takes the statistical memory about previous states of the system
into account. By using Gram-Schmidt orthogonalization procedure
\cite{Yulm1} we receive the recurrent formula, in which the senior
dynamic variable $ {\bf W} _n = {\bf W} _n (t) $ is connected to
the junior one in the following way: \be { \bf W} _0 = {\bf A}_k^0
(0), ~~{\bf W}_1 = \{ i \hat L-\lambda_1 \}{\bf W}_0, \ldots ~~
{\bf W}_n = \{ i\hat L-\lambda _ {n-1} \} {\bf W}_{n-1}+
\Lambda_{n-1} {\bf W}_ {n-2}+..., ~~ n > 1. \label {f11} \ee
Introducing the corresponding projection operators we come to the
following  chain of connected non-Markov finite-difference kinetic
equations ($t=m\tau, n=1,2,... $) \be \frac {\Delta M_n (t)}
{\Delta t} = \lambda _ {n+1} M_n (t) -\tau
\Lambda _ {n+1} \sum _ {j=0} ^ {m-1} M _ {n+1} (j\tau) M_n (t-j\tau). \\
\label {f12} \ee Here $ \lambda _ {n+1} $ is an eigen-value of the
Liouville's quasioperator, and relaxation parameters $ \Lambda _
{n+1} $ are determined as follows \bn \lambda_n=i \frac {< {\bf W}
_n \hat L {\bf W} _n >} {< | {\bf W} _n | ^ 2 >}, \ \Lambda_n
=-\frac {< {\bf W} _ {n-1} (i \hat L - \lambda _ n) {\bf W} _n >}
{< | {\bf W} _ {n-1} | ^2 >}, ... \ . \nonumber \en

In this work we analyze the short time series with the help of
Eqs. \Ref {f8}-\Ref {f12} where non-stationary functions may not
be taken into account \cite {Yulm3}. So, we shall use full set of
the dynamic, kinetic and spectral parameters, functions and
characteristics as a quantitative information measure to describe
of spread of epidemic in human population. Among them, here are
temporal dynamics of the first four orthogonal dynamic variables,
phase portraits in some planes of dynamic variables, TCF, junior
memory functions and their power spectra, frequency dependence of
first three points of non-Markovity parameter, and local time
behavior of the locally average relaxation parameters.

\section{Experimental data and data processing}
The registration and accounting materials, as well as the results
of certain studies are the initial data for epidemic research in
medicine. We have made use of the accounting materials of Kazan
sanitary - epidemiological stations located in industrial
districts. The concrete data were given by the sanitary -
epidemiological station of the Privolzhskii district of Kazan
covering the period from 10.27.1995 to 03.05.2002. Figs. 1.A
presents the experimental raw data of epidemic spread, involving
the total population of the Privolzhskii district (about 300
thousand people). The first group of the data represents a
six-year dynamics of grippe, the second group describes a six-year
dynamics of acute respiratory track infections in this district.
The obtained data were processed with the help of the above
introduced technique. The set of three memory functions was
calculated for each sequence of the data. The power spectra for
each of these functions are obtained by the fast Fourier transform
(FFT). We will also show the phase portraits in plane projections
of the multidimensional space of the dynamic orthogonal variables.
For a more detailed diagnosis of the system we will consider the
frequency spectrum of the first three points of the statistical
spectrum of the non-Markovity parameter. In this study we will use
the frequency dependence of the statistical spectrum of
non-Markovity parameter \cite{Yulm1}-\cite{Yulm3} \be
\varepsilon_i (\omega) = \left\{ \frac {\mu _ {i-1} (\omega)} {
\mu_i (\omega)} \right\}^{\frac {1} {2}}. \label{f13} \ee Here
$i=1, 2.. $ and $ \mu_i (\omega) $ is a power spectrum for $i$th
order memory function of the $i$th relaxation level. The
statistical spectrum of non-Markovity parameter $\varepsilon_i
(\omega)$ constitutes an information measure of Markovity and
non-Markovity or randomness and regularity in time evolution of
underlying systems.

\section{Discussion of the results}
In this section the quantitative and comparative analysis of the
six-year period of grippe chaotic dynamics (Fig. 1.B.a) and of
ARTI (Fig. 1.B.e) will be carried out on the basis of the theory
submitted in Section 3. The time series of the initial signals,
the phase portraits of the dynamic variables, the power spectra of
TCF and the junior functions of memory as well as frequency
dependence of the first three points of the statistical non-Markov
parameter are submitted in Figs. 1-4. We develop a new approach in
the study of epidemic processes on the basis the local time
behavior of the relaxation and kinetic parameters $\lambda_1$,
$\lambda_2$, $\lambda_3$, $\Lambda_1$ and $\Lambda_2$ (Figs. 5, 6)
also.

The six-year sampling of grippe  and ARTI dynamics are submitted
in Fig. 1.B. (Figs. 1.B. a-d for grippe, Figs. 1.B. e-h for ARTI).
The time series of the orthogonal variable $W_0 $ (Fig. 1.B.a),
$W_1 $ (Fig. 1.B.b), $W_2 $ (Fig. 1.B.c), $W_3 $ (Fig. 1.B.d) for
the chaotic dynamics of grippe and ARTI have appreciable symmetry
relative to straight line $W_i=0 $, where $i = 1...3$. The
analysis of the time series of variable $W_0 $ (Fig. 1.B.a) for a
grippe  epidemic shows that the moment of the grippe outburst or
appearance of epidemic corresponds to the most significant
fluctuations of this dynamic variable. Epidemic outbursts of the
disease are located at equal distances from each other. This is
the evidence of periodicity and interval constancy between grippe
epidemics within the whole time of observation. The time series
for three orthogonal variables $W_1 $ (Fig. 1.B.b), $W_2 $ (Fig.
1.B.c), $W_3 $ (Fig. 1.B.d) of grippe epidemics are almost
perfectly symmetrical. The time series of dynamic variables $W_0 $
(Fig. 1.B.e), $W_1 $ (Fig. 1.B.f), $W_2 $ (Fig. 1.B.g), $W_3 $
(Fig. 1.B.h) of ARTI epidemiological processes present a different
picture. The visible fluctuations of the dynamic variables for
ARTI are observed in the autumn-winter period same as in case of
grippe epidemics. However, dynamic noises can be observed between
the outbursts of ARTI epidemics. They are represented by
monotonously inequable parabolas. One of the half-parabola
corresponds to one of the epidemics and another corresponds to the
subsequent one. The time development of orthogonal variables $W_1
$ (Fig. 1.B.f), $W_2 $ (Fig. 1.B.g), $W_3 $ (Fig. 1.B.h) of the
ARTI epidemiological process represents symmetric series about
straight line $W_i=0 $. There are seven outbursts in the time
behaviour of the dynamic variables in the seven autumn - winter
periods (from 10.27.1995 to 03.05.02). They corresponds to the
seven cycles of grippe and ARTI epidemics. However the time series
of orthogonal variables $W_i $ ($ i=1,2,3 $) for ARTI differ from
grippe's case by the strongly expressed asymmetry and  the
existence of a clearly expressed noise.

In Fig. 2 the phase clouds in the six plane projections of the
four first dynamic variables $W_i $, $i=0 ... 3 $ for epidemic
processes of grippe  and ARTI are submitted. In case of grippe the
phase clouds have a well defined asymmetry about the centre of
coordinates. All the phase clouds contain a centralized nucleus
and an aggregate of points scattered on perimeter in a fan-shaped
way. The interval of the dispersal makes up 900 $\tau$. Owing to
such asymmetric stratification the phase clouds get an elongated
shape. The phase portraits of the ARTI epidemiological process
have smaller symmetry about the centre of coordinates. The nucleus
of the portrait gets the shape of a right angle and resembles a
"boot". Apical emissions of separate points are appreciable on
each side of the "boot". The interval of the dispersal of the
points decreases up to 700 $\tau$. The stratification of the phase
clouds of a grippe epidemiological process is more essential, than
in case of an ARTI epidemic. The analysis of the data in other
Kazan districts reveals a prominent phase portrait of a grippe
epidemic. It is the elongation of the phase portrait for an ARTI
epidemic namely its "boot-type" shape.

In Fig. 3 the power spectra of TCF $ \mu_0 (\omega) $ and three
junior memory functions  $ \mu_i (\omega), i=1, 2, 3 $ for grippe
and ARTI epidemiological processes are submitted. The frequency
spectra of $ \mu_0 (\omega) $ are given in doubly-log scale for a
more detailed analysis of the data.  In case of grippe the fractal
dependence in the region of intermediate  and high frequencies is
found in the power spectrum of the initial TCF $ \mu_0 (\omega) $.
The area of these frequencies is divided by a small outburst of
power on frequency $\omega = 3*10^{-1} f.u.\ (1 f.u.=2 \pi/\tau)$.
In the domain of the low frequencies sharp breaks of the spectrum
are observed. The frequency dependency of the next three junior
memory functions $ \mu_i (\omega), i=1, 2, 3 $ are accompanied by
spectral outbursts at equal intervals and by condensation of
spectral lines near the outbursts. A spectral noise have been
presented in all diagrams of the frequency dependencies. The
spectrum of the initial TCF $ \mu_0 (\omega) $ for an ARTI
epidemic process differs from a grippe epidemic only in the high
frequency area. Additional spectral peaks are appreciable at
frequencies $ \omega = 10 ^ {-1} f.u. $ and $ \omega = 10 ^ {-1/2}
f.u.$ The power spectra of the three junior memory functions $
\mu_i (\omega), i=1, 2, 3 $ for an ARTI epidemic have bright
fractal dependence with several spectral peaks at every $0.15
f.u$.

In Fig. 4 the spectra of the first three points of statistical
non-Markov parameter $ \varepsilon_i (\omega) $, where $i=1,2,3 $
are presented. Fractalness is appreciable only in the behaviour of
$\varepsilon_1 (\omega) $. With approximation to frequency $
\omega = 0 $ the monotonous amplification of markovization is
observed. The spectra of the non-Markovity parameter for the
second point of $ \varepsilon_2 (\omega) $ for ARTI and grippe
epidemics accept value $ \varepsilon_2 (\omega) = 1$  are almost
identical but it have small fluctuations. It testifies to strong
non-Markovity. The sinuosity with two frequency crests in area $
\omega =  0.1$ $ f.u. $ and $ \omega = 0.3 $ $ f.u. $ is
appreciable in frequency spectra of the third point of statistical
non-Markovity parameter $ \varepsilon_3 (\omega)$.

In Fig. 5 the time dependence of local (with local averaging)
kinetic and relaxation parameters $ \lambda_1 $, $ \lambda_2 $, $
\lambda_3 $, $ \Lambda_1 $ and $ \Lambda_2 $ for a grippe epidemic
is submitted. The sampling for the local averaging includes 200
points. The similar size of the sampling allows to consider the
physical nature of a real object (a grippe and ARTI epidemics) in
a more detailed way. Breaks are observed at a great number of
points in the time scanning of the parameters. In the time
dependence of all the parameters within six years of every autumn
- winter period the system drops out of the steady state and
returns to it. The whole process can be presented as recurrence of
two various processes. At first small fluctuations of the
parameter are observed. The fluctuations visibly grow in the
moment of seasonal epidemics. Then the system acquires the state
of quasistable balance. The time interval for each process is 125
points, that corresponds precisely to one of the autumn - winter
periods.

Recently the correlation analysis has experienced a marked lack of
information concerning the object of the research.  Procedure of
local averaging of various parameters allows to examine the
separate hidden  properties of studied objects. The characteristic
feature of the usual correlation analysis is the fact that the
greatest possible set of signals is required for the qualitative
analysis of the properties of the object of the research. At
longer sample of such signals it is possible to receive more exact
information with the help of the correlation analysis.  Let us
take a random non-Markov process as an example. This process
consists of a sequence of alternating states. Thus the problem of
extraction more information not only about the common process but
about various separate states of the system arises. In this case
the use of the correlation analysis for all the time series will
be inefficient. The processing of the signals is necessary  for
separate local sites of the whole time series. It will allow to
consider the properties of separate dynamic states of the system.

Hereinafter a new method of data processing based on the local
averaging of kinetic and relaxation parameters is offered. This
method allows to consider the properties of separate,
non-stationary states of the systems. The idea of a method  is the
following: there exists an initial data set. Let's take a sampling
in length N of signals and calculate its kinetic and relaxation
parameters. Then the operation of "step-by-step shift to the
right" for one time interval is carried out. The kinetic and
relaxation parameters are calculated again. The "step-by-step
shift to the right" is continued to the end of a time series. Such
locally averaged parameters have high sensitivity to the effects
of intermittency and nonstationarity. If the initial time series
has some irregularity, it is instantly reflected in the behavior
of the locally averaged parameters.

The use of this method requires the choice of the optimal length
of a sampling which enables to receive the most trustworthy
information. If a sampling is too short, noise effects do not
allow to receive qualitative information. Besides with a short
length sampling we have significant errors. On the other hand at a
great length of a sampling locally averaged parameters lose
"sensitivity" necessary for the study. As a result of the studies
of different lengths of local sampling we have received the
optimal length which makes 100-120 points. Further proofs of all
aforesaid will be given below.

In Fig. 6 the time dependence of local kinetic and relaxation
parameters $ \lambda_1 $, $ \lambda_2 $, $ \lambda_3 $, $
\Lambda_1 $ and $ \Lambda_2 $ for an ARTI epidemic is submitted.
200 points where selected for the procedure of localization. Due
to a more significant noise of the ARTI chaotic dynamics the time
scanning of then parameters essentially differs from  grippe.  The
time dependence of the first parameter $ \lambda_1 $ reflects the
physical features of an ARTI epidemic process. Instead of small
fluctuations at the beginning of the epidemic sharp fluctuation
splashes appear. Between them the system acquires a rather quiet
state followed by a new epidemic peak. When the epidemic is over
only small fluctuations in the system can be appreciable. This
process  repeats periodically during all long-term interval. Some
definite regularity can be observed in the time dependence of the
next kinetic and relaxation parameters. The sharp peak at the
beginning and at the end of the ARTI epidemic process relates to
them. Only small fluctuations of the system are appreciable at the
time of epidemic. After each ARTI epidemic the considered system
comes back to the state of relative balance.

Judging by the behavior of the parameters the distinction between
the epidemic of grippe  (I) and that of ARTI (II) comes to the
following. The initial values of the intensity $ \mu_i (\omega =
0) $ for all $ i = 0, 1, 2, 3 $  for systems I and II  differ by
more than one order. It is the evidence of remarkable distinction
in times of relaxation (correlations). For system I these times
are almost 20-50 times shorter, than for system II. The intensity
scales also differ almost by 50 times. Thus the initial values of
$\varepsilon _ {1} (\omega=0)$ differ almost by 5 times for
systems I and II.

On the basis of the aforesaid it is possible to come to certain
conclusions about a great randomness of grippe dynamics in
comparison with ARTI dynamics.  The last one has greater
regularity and predictability. Other feature of the investigated
epidemics is the existence of the three additional well - defined
spectral splashes except the basic annual. From figs. 3 b-d and 3
f-h one can see that these groups of spectral peaks correspond to
the characteristic frequencies equal accordingly to 0.14 $ f.u.,
0.28 f.u $. and 0.43 $ f.u.$ The intensity of  the high-frequency
peak is almost an order lower, than the two other peaks of lower
frequencies.

The strong distinction in the dynamic behavior of  grippe  and
ARTI epidemics is shown in local time dependence of relaxation
constants $\lambda_1$ , $\lambda_2$, $\lambda_3$, $\Lambda_1$ and
$\Lambda_2$. It is necessary to remind, that parameters $
\lambda_i, i=1,2,3... $ remind Luapunov's exponents for various
relaxation levels. For system II great regularity is observed. All
parameters of $ \lambda_1 $, $ \lambda_2 $, $ \lambda_3 $ are
negative, which is the evidence of the relative stability of the
system. For system I (grippe) owing to the general instability the
parameter $ \lambda_1 (t) $ gets small positive values on tail. It
corresponds to the slump of epidemic. From Eq. (8) one can see,
the system becomes unstable at values $ \lambda_1 (t)> 0 $.

The time behavior of parameters $\Lambda_1(t)$ and $\Lambda_2(t)$
is more dramatic. If  $ \Lambda_i (t) > 0, i=1,2 $ then solution
of Eq. (8) corresponds to a steady condition of social network, if
$ \Lambda_i (t) < 0 $ instability grows. From Figs. 5 d, e and 6
d, e it is clear, that annual epidemics of grippe have a greater
instability of the system. In case of ARTI the behavior of $
\Lambda_2 (t) $ everywhere has steady state of system II. However
the parameter of $ \Lambda_1 (t) $ in the greater part of the time
scale is negative, this testifies to a loss of stability in this
area.

So, according to our theory, parameter $\lambda_1(t)$= $R$ make
sense a relaxation rate  of random process. In case of grippe (I)
isoline exists in behavior $\lambda_1$=$\lambda_1(t)$ with values
0.08 $\tau^{-1}\le R \le$ 0.24$\tau^{-1}$. Simultaneously, a
relaxation rate on recession of grippe epidemic sharply grows up
to (0.8-1.0)$\tau^{-1}$, that it is in 4 - 12 times,
approximately. This implies the greater stabilization of system,
than during rise or on a peak of epidemic (see, fig. 7 a, b for
more details).

In case ARTI isoline in behavior $\lambda_1(t)$ does not exist at
all. It testifies about quasistability and seasonal prevalence in
relaxation  behavior, in - whole. Absolute values of relaxation
rates fluctuate in an interval 0.24 $\tau^{-1}\le R \le$
0.6$\tau^{-1}$. On a tail a relaxation rate on recession of
epidemic ARTI speed of grows in 2-2.5 times. Therefore system, as
a whole, has the greater macroscopic stability.

We can draw the following conclusion from comparison of relaxation
parameter $\lambda_1(t)$ of grippe and ARTI (see, table 1 for the
whole sampling). Relaxation rates differ almost in 3.4 times. It
testifies to macroscopic stability of epidemiological process of
ARTI. On the other hand, local amplification  of relaxation rate
(on a tail of  epidemic) in case of a grippe creates an additional
source of stabilization for unstable,  in the  whole, process of
grippe epidemic.

\section{Conclusions}
In this paper we have shown that the chaotic dynamics of epidemic
processes of grippe  and ARTI in real social network can be
considered as a stochastic discrete non-Markov process. It is more
convenient to use the statistical theory of the discrete
non-Markov processes for real objects and alive systems \cite
{Yulm1} - \cite {Yulm3} to study the similar  processes. The used
theory allows to define essential distinctions between the
epidemic processes of grippe and acute respiratory track
infections by degrees of randomness, regularity and
predictability. By processing  the experimental data about
patient's numbers we received the time series of dynamic variables
$W_i (t) $  and calculated the memory functions and the first
three points of the statistical spectra of non-Markovity
parameter. We used the power spectra, which were received with
help of the fast Fourier transform (FFT) to analyze the diverse
time functions (correlation and memory functions).

The new qualitative approach in studying statistical properties of
epidemic processes of grippe  and ARTI originates while using
locally average kinetic and relaxation parameters $\lambda_1$,
$\lambda_2$, $\lambda_3$, $\Lambda_1$ and $\Lambda_2$.

It is possible to come to the following conclusions on the basis
of the conceptions about long-range memory and localization of
parameters. The epidemic process of grippe causes a great danger
to mankind because of its randomness and unpredictability. Each
epidemic process of grippe is characterized by a sudden outburst
and sudden attenuation. The peak of an epidemic lasts a
comparatively short period of time (60 days). It means efforts to
liquidate the consequences of a grippe epidemic should be made
during this period of time. The epidemic process of ARTI is
characterized by a greater regularity and predictability. It is
accompanied by a small outburst at the beginning of the epidemic,
then the number of patients diminishes. After it a drastic
outburst corresponding to the epidemic peak is observed. The
process ends with small short-term splashes which later quickly
fade. Unlike grippe, an ARTI epidemic is characterized by a less
drastic outburst and attenuation. During the underlying interval
insignificant outbursts of respiratory infections are observed
followed  by periodic outbursts of ARTI.

We have revealed and explained essential distinctions in phase
portraits, frequency spectra and quantitative estimations of
memory functions and statistical non-Markov parameters between the
epidemic processes of grippe and ARTI. Relative distinctions of
the quantitative data about epidemics of grippe  and ARTI are the
evidence of the distinction in long-range memory, order and
organization of each process in the certain social system. These
distinctions allow to define the degree of importance of each of
the epidemics. In particular, we have shown that the epidemic
process of grippe  and respiratory infections  have essentially
different quantitative measure of randomness and regularity.

The local kinetic and relaxation parameters of epidemic processes
of grippe  and ARTI allow to study statistical features of these
systems in detail. The local time dependencies allow to find the
internal features of the epidemic process. It  helps in the study
of real objects. With abundant experimental data the developed
method allows to define the laws of epidemic processes.

The received results can be of practical value while studying
other diseases and comparative estimation of epidemic danger.

In this paper we have demonstrated that the set of relaxation,
kinetic and spectral parameters and characteristics of discrete
non-Markov stochastic processes are valuable for the  description
of the role of randomness,  regularity and predictability of
epidemic processes, the spread of grippe and ARTI in the world.

Since the similar situation is typical of the majority of epidemic
diseases on  networks our conclusions are of profound importance
for a large number of physical, biological and technological
networks.The results received here are connected with
non-stationary and non-ergodic processes in stochastic systems.
Therefore they are certainly of value for physics of the disorder
matter in which similar processes can occur on molecular or
structural levels.

Distribution of atypical pneumonia (SARS) attaches special value
to any researches of ARTI. The cases similar SARS, unfortunately
can  a  multiply in the future . In connection with stated  above
a study of  ARTI  infections represents the important scientific
significance.

\section{Acknowledgements}
This work  supported by the  RHSF (Grant N 03-06-00218a), RFBR
(Grant N 02-02-16146, 03-02-06084, 03-02-96250) and CCBR of
Ministry of Education  RF (Grant N E 02-3.1-538). The authors
acknowledge Dr. L.O. Svirina for technical assistance.

\section{Figures Caption}
FIG. 1.A. Six-year temporal dynamics of the experimental raw data
of grippe (1.A.a) and acute respiratory track infections (1.A.b)
in Privolzhskii district of Kazan, Russia, from 10.27.95 to
03.05.02. The dynamics of ARTI is initially characterized by a
slow increase and then by slow decrease in the number of patients.
With grippe we have a more sharp picture, characterized by low
level of the background noise, then by a sudden increase and
decrease of the disease.

FIG. 1.B. Time series of orthogonal variables $W_0 $ (a, e), $W_1
$ (b, f), $W_2 $ (c, g), $W_3 $ (d, h) of chaotic dynamics of
grippe (a, b, c and d) and acute respiratory track infections
(ARTI)(e,f,g and h). The spectral peaks in time series display the
seasonal periodic occurrence of grippe and ARTI epidemics. The
time series of orthogonal variables for the epidemic process of
ARTI differ due to the occurrence of an appreciable noise.

FIG. 2.  The phase clouds for epidemic processes of grippe and
acute respiratory track infections (ARTI) in plane projections for
the various combinations of orthogonal variables $W_i $, $W_j $
where $i, j=0,1...3 $. During the epidemic of grippe  the phase
points disperse about the nucleus. The phase clouds for the
epidemic process of ARTI acquires the shape of a "boot" and are
grouped about nucleus.

FIG. 3. The power spectra of time correlation function (TCF) $
\mu_0 (\omega) (a) $ and of the three junior memory functions $
\mu_i (\omega), i=1 (b), 2 (c), 3 (d) $ for the epidemic processes
of grippe and acute respiratory track infections (ARTI). For a
more detailed analysis the power spectrum of TCF $ \mu_0 (\omega)
(a) $ is submitted in double logarithmic scale. In the field of
high frequencies three additional spectral splashes are observed.
The spectral peaks in the registered spectra are the evidence of
of long-range memory in the system.

FIG. 4. The frequency dependence of the first three points of
statistical non-Markov parameter $ \varepsilon_i (\omega) $ for
epidemic processes of grippe and acute respiratory track
infections (ARTI). For the first point of $ \varepsilon_1 (\omega)
$ the splash of markovization process is observe in the vicinity
of value $ \omega = 0 $. The values of other parameters $
\varepsilon_2 (\omega) $ and $ \varepsilon_3 (\omega) $ change in
a narrow interval about a unit. This reflects the existence of
long-range memory order in the system.

FIG. 5. The time dependence of local kinetic and relaxation
parameters $ \lambda_1 $, $ \lambda_2 $, $ \lambda_3 $, $
\Lambda_1 $ and $ \Lambda_2 $ for  the epidemic process of grippe
in units of $\tau$. The sampling of local averaging is equal to
200 points. The sudden dropout of the system from a  balanced
condition is determined by the epidemic of grippe. The reversion
of the system into a former steady state defines the intervals
between the epidemics of grippe.

FIG. 6. The time dependence of local kinetic and relaxation
parameters $ \lambda_1 $, $ \lambda_2 $, $ \lambda_3 $, $
\Lambda_1 $ and $ \Lambda_2 $ for epidemic process of acute
respiratory track infections (ARTI) in units of $\tau$. The length
of the sampling of local averaging is equal to 200 points. The
sharp splash in a time series demonstrates the beginning of an
ARTI epidemic. The end of the epidemic process is defined by the
following peak.  At the moment of epidemic the system is in an
unstable state.

FIG. 7. Time dependence of local relaxation rate $\lambda_1 (t)$
for grippe (b) and ARTI (d) with initial row data for grippe (a)
and ARTI (c). We show extracted sampling for comparison. One can
see the dramatic revising of $\lambda_1(t)$ during both epidemic.
Time behavior $\lambda_1(t)$ at grippe contains isoline's regime
whereas  one can notice continuous alteration of $\lambda_1(t)$ in
case of ARTI.

\begin{figure}[htbp]
\includegraphics[angle=0]{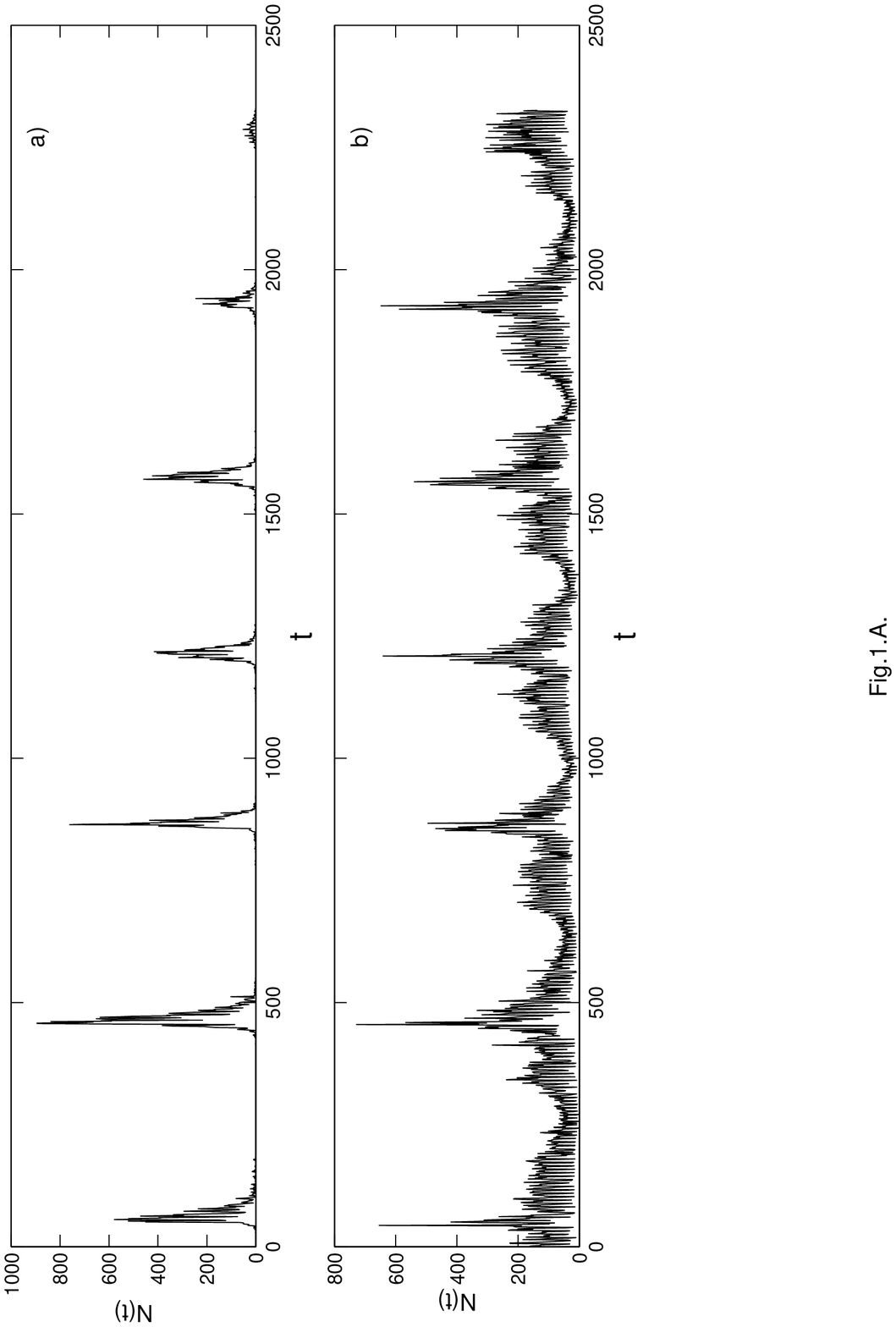}
\caption{}
\end{figure}

\begin{figure}[htbp]
\includegraphics[angle=0]{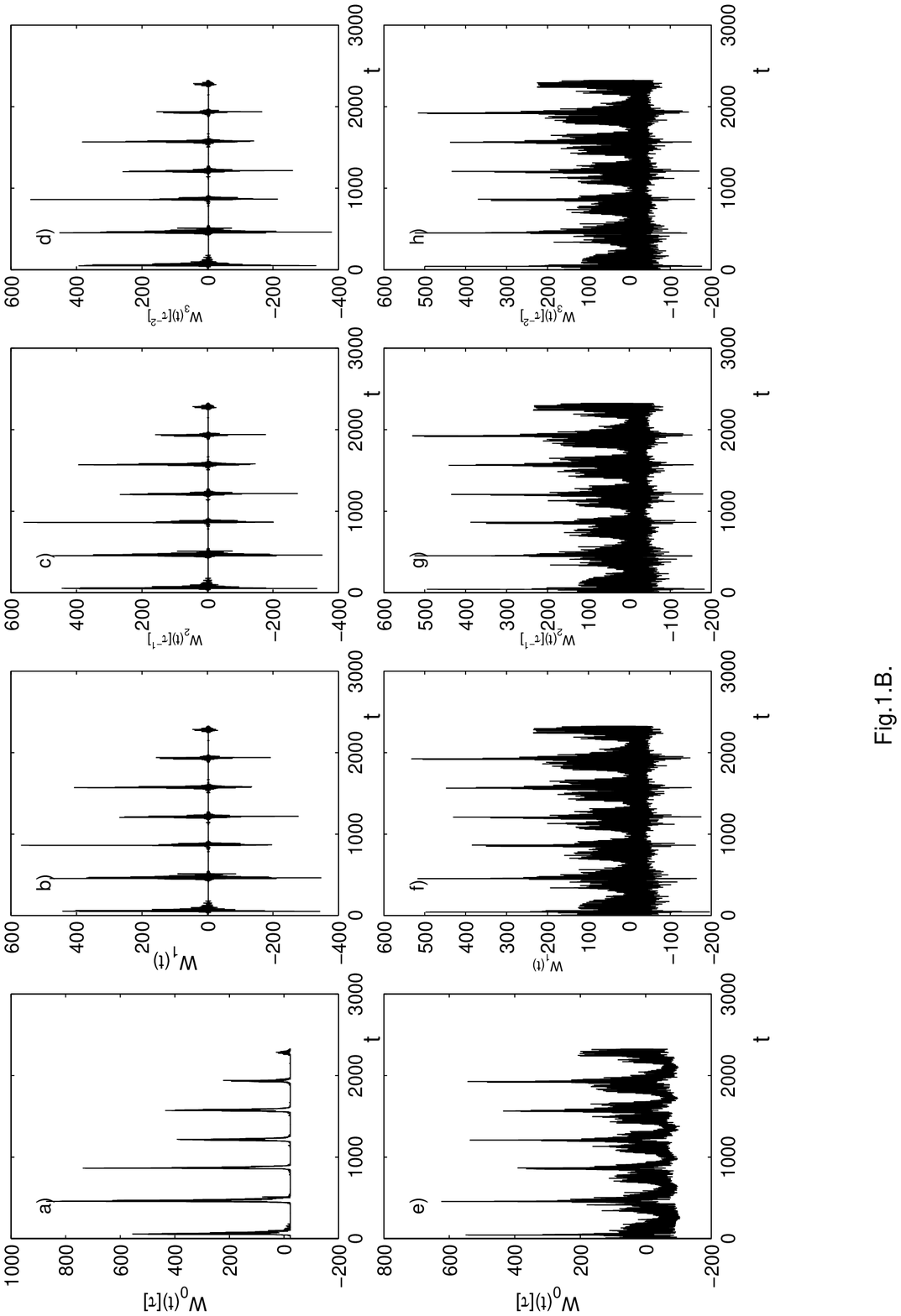}
\caption{}
\end{figure}

\begin{figure}[htbp]
\includegraphics[angle=0]{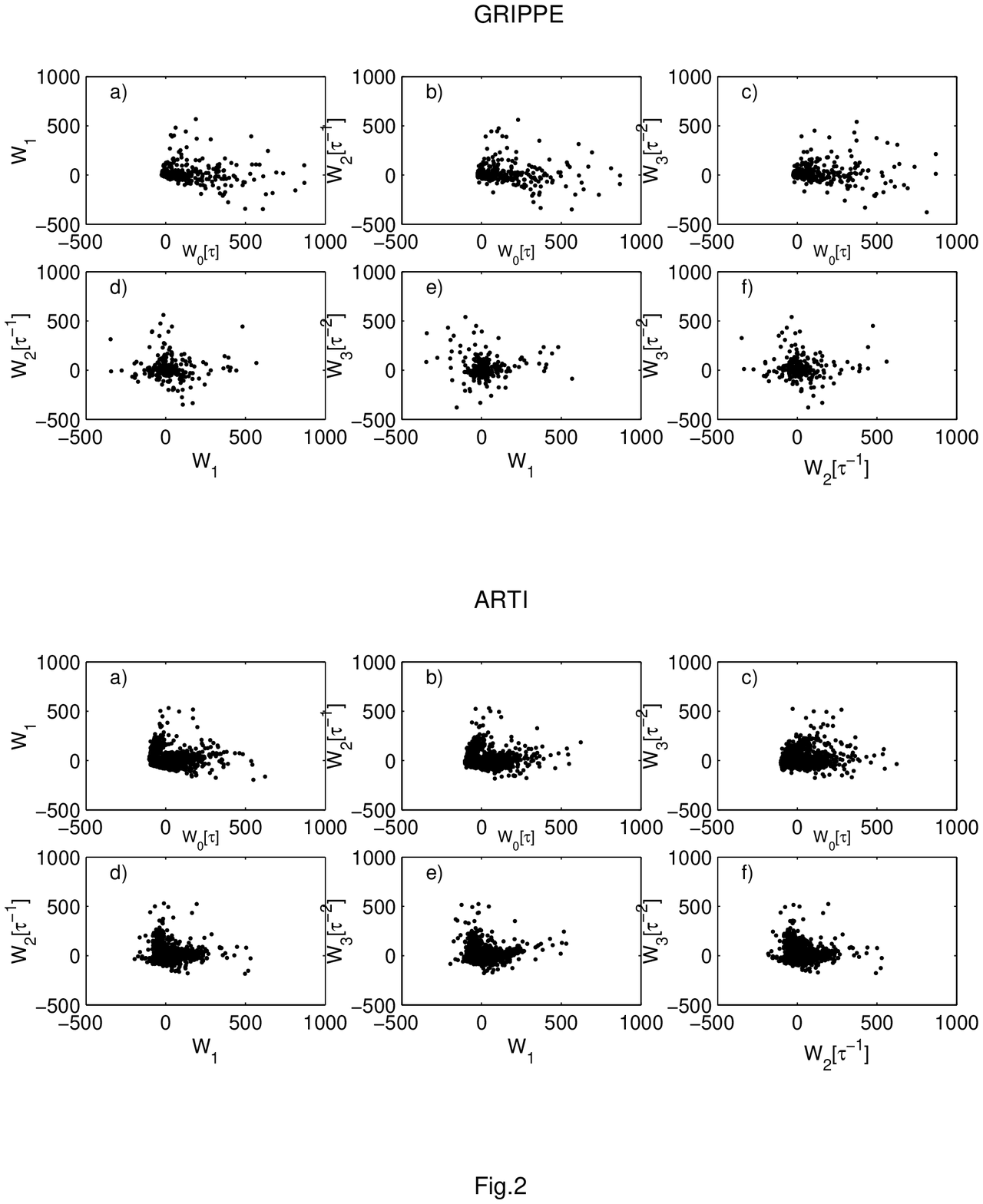}
\caption{}
\end{figure}

\begin{figure}[htbp]
\includegraphics[angle=0]{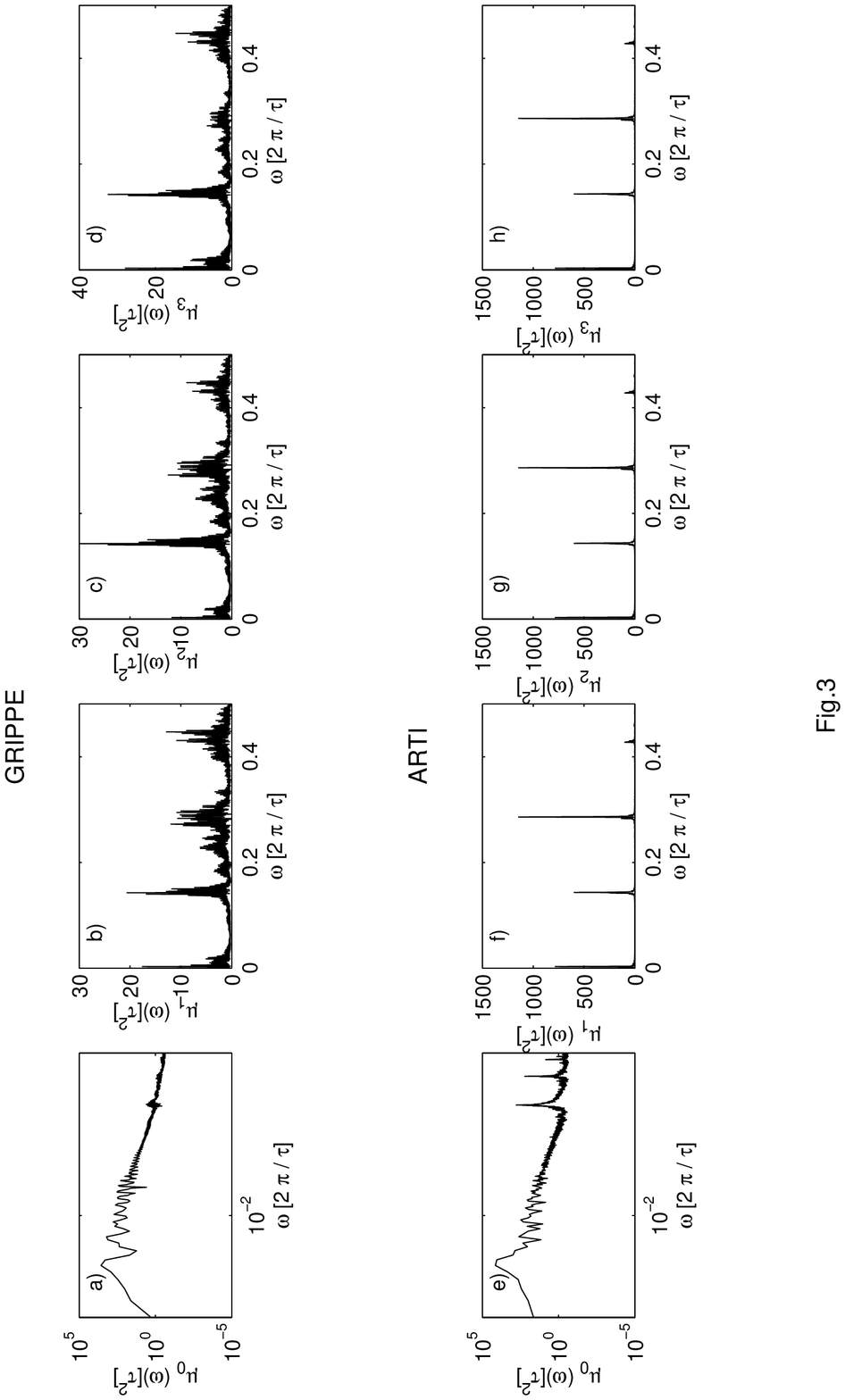}
\caption{}
\end{figure}

\begin{figure}[htbp]
\includegraphics[angle=0]{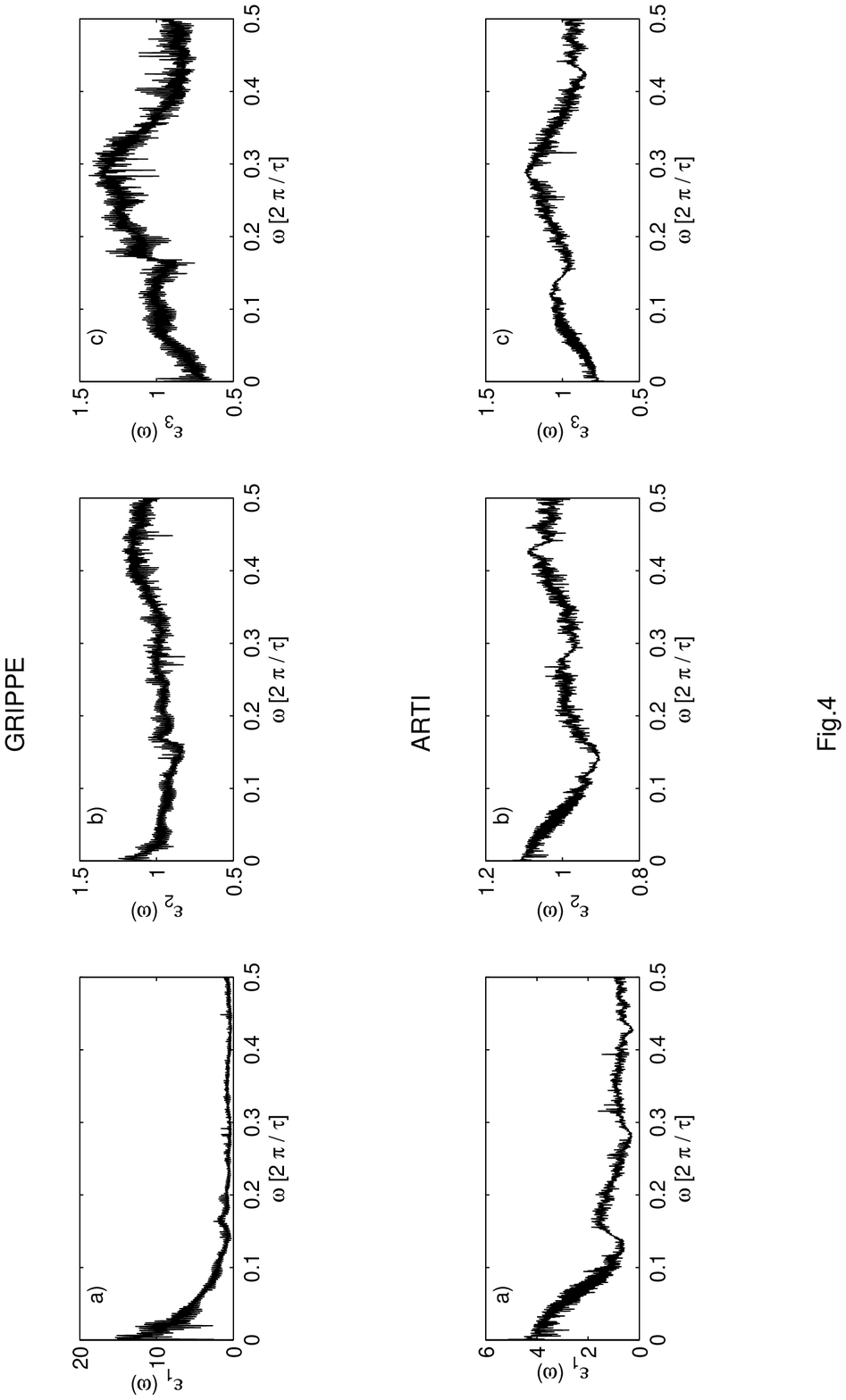}
\caption{}
\end{figure}

\begin{figure}[htbp]
\includegraphics[angle=0]{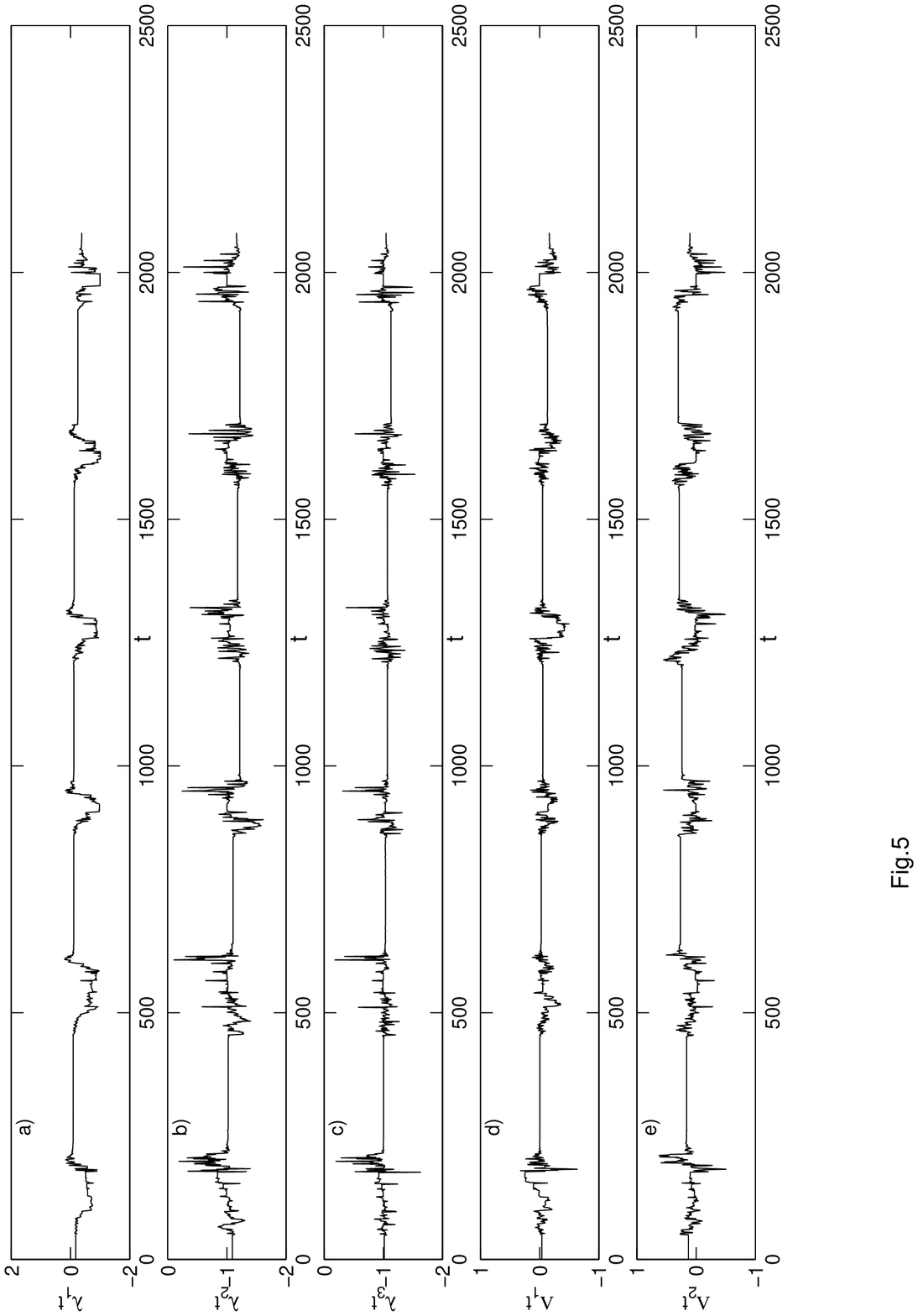}
\caption{}
\end{figure}

\begin{figure}[htbp]
\includegraphics[angle=0]{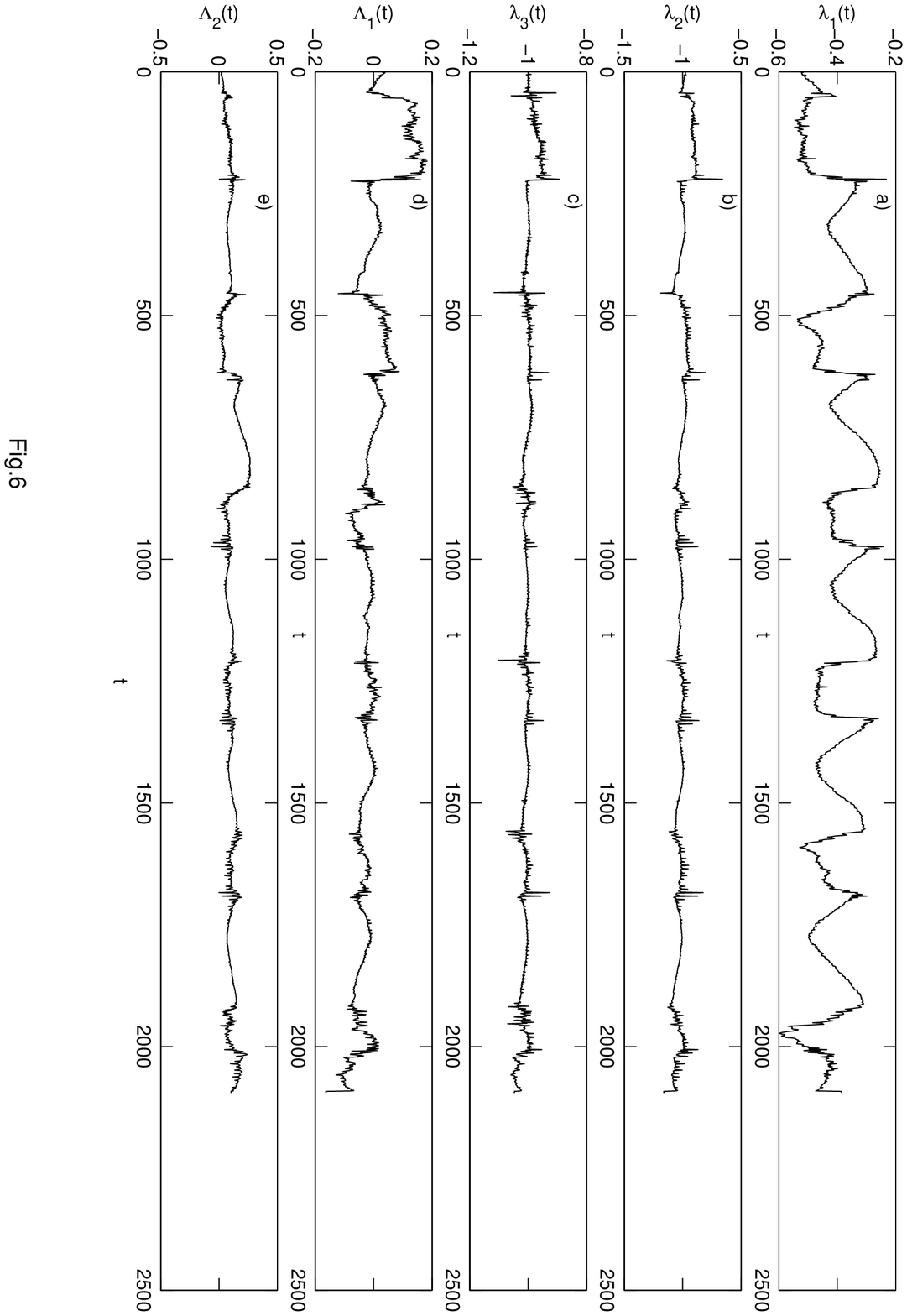}
\caption{}
\end{figure}

\begin{figure}[htbp]
\includegraphics[angle=0]{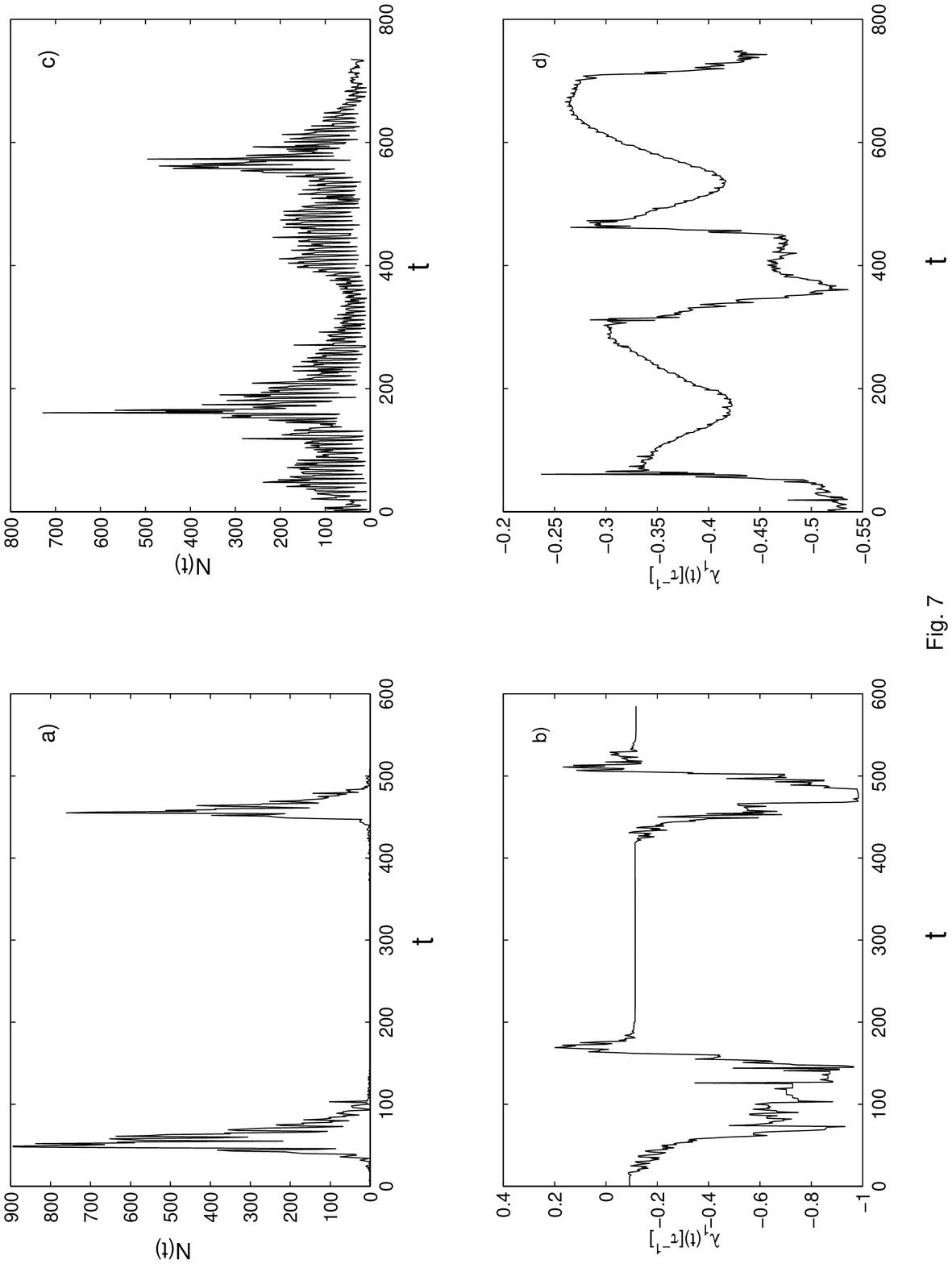}
\caption{}
\end{figure}

\end{document}